\documentclass[reprint,aps,amsmath,amssymb,twoside,prd,showkeys,superscriptaddress]{revtex4-1}
\pdfoutput=1
\usepackage{verbatim}
\usepackage{comment}
\usepackage{graphicx}
\usepackage[T1]{fontenc}
\usepackage{epsfig}
\usepackage{bm}
\usepackage{tensor}
\usepackage{amssymb}
\usepackage{float}
\usepackage{amsmath}
\usepackage{subfigure}
\usepackage{dcolumn}
\usepackage[utf8]{inputenc}
\usepackage{cancel}
\usepackage[colorlinks]{hyperref}
\usepackage[usenames,dvipsnames]{color}
\hypersetup{
     breaklinks=true,
    pdfstartview={FitH},    
    colorlinks=true,       
    linkcolor=blue,          
    citecolor=red,        
    filecolor=magenta,      
    urlcolor=blue,           
    anchorcolor=green,      
    linktocpage=true
}

\def\doi{http://doi.org}

\newcommand{\onehalf}{{\textstyle\frac{1}{2}}}

\newcommand{\onethird}{{\textstyle\frac{1}{3}}}
\newcommand{\quarter}{{\textstyle\frac{1}{4}}}
\newcommand{\threequarter}{{\textstyle\frac{3}{4}}}

\newcommand{\eref}[1]{Eq.~(\ref{#1})}

\newcommand{\HCd}{\mathcal{H}}

\def\HCdt0{\tilde{\HCd}_{0}}

\newcommand{\afffias}{Frankfurt Institute for Advanced Studies (FIAS), 
Ruth-Moufang-Strasse~1, 60438 Frankfurt am Main, Germany}
\newcommand{\affbgu}{Physics Department, Ben-Gurion University of the Negev, 
Beer-Sheva 
84105, Israel}

\newcommand{\affguni}{Fachbereich Physik, Goethe-Universitat, Max-von-Laue-Strasse 1, 60438 Frankfurt am Main, Germany}

\begin{document}
\title{Low-Redshift Constraints on Covariant Canonical Gauge Theory of Gravity}
\author{David Benisty}
\email{benidav@post.bgu.ac.il}
\affiliation{\afffias}\affiliation{\affbgu}
\author{David Vasak}
\email{vasak@fias.uni-frankfurt.de}
\affiliation{\afffias}
\author{Johannes Kirsch}
\email{jkirsch@fias.uni-frankfurt.de}
\affiliation{\afffias}
\author{Jürgen Struckmeier}
\email{struckmeier@fias.uni-frankfurt.de}
\affiliation{\afffias}\affiliation{\affguni}
\begin{abstract}
Constraints on the Covariant Canonical Gauge Gravity (CCGG) theory from low-redshift cosmology are studied. The formulation extends Einstein's theory of General Relativity (GR) by a quadratic Riemann-Cartan term in the Lagrangian, controlled by a "deformation" parameter. In the Friedman universe this leads to an additional geometrical stress energy and promotes, due to the necessary presence of torsion, the cosmological constant to a time-dependent function. The MCMC analysis of the combined data sets of Type Ia Supernovae, Cosmic Chronometers and Baryon Acoustic Oscillations yields a fit that is well comparable with the $\Lambda$CDM results. The modifications implied in the CCGG approach turn out to be subdominant in the low-redshift cosmology. However, a non-zero spatial curvature and deformation parameter are shown to be consistent with observations. 
\end{abstract}
\maketitle

\section{Introduction}
\label{sec:Intro}
Dark energy, inflation, and dark matter are examples of much disputed concepts that have been added to Einstein's General Relativity (GR) in order to explain observations that otherwise would not be accounted for, see e.g. \cite{Weinberg:1988cp,Lombriser:2019jia,Frieman:2008sn}. The cosmological constant, as its value adjusted to fit the current accelerated expansion of the universe is far at odds with the calculated vacuum energy of matter which it is supposed to represent. Quintessence \cite{Wetterich:2014gaa,Ratra:1987rm,Caldwell:1997ii,Kehayias:2019gir,Oikonomou:2019muq,Chakraborty:2019swx,Babichev:2018twg,Zlatev:1998tr,Caldwell:1999ew,Chiba:1999ka,Bento:2002ps,Tsujikawa:2013fta} -- and similar scalar fields invoked to generate an initial explosive inflation of the universe and explain the measured isotropy of cosmic radiation -- lack fundamental physical underpinning. Modifications of the gravity are, among other models, hand-crafted just for matching specific observations. The invisible dark matter, finally, necessary to explain the dynamics of galaxies, could not yet been attributed to any field theory, or a known or unknown particle, despite astronomical budgets devoted to its search \cite{DiValentino:2020vhf,DiValentino:2020zio,DiValentino:2020vvd,Efstathiou:2020wxn,Borhanian:2020vyr,Hryczuk:2020jhi,Klypin:2020tud,Ivanov:2020mfr,Chudaykin:2020acu,Lyu:2020lps,Alestas:2020mvb,Motloch:2019gux,Frusciante:2019puu,Yang:2020uga,DiValentino:2020vnx,DiValentino:2020leo,Benaoum:2020qsi,Yang:2020myd,DiValentino:2020kha,DiValentino:2020evt,Yang:2020tax}.

Recently a novel, rigorously derived covariant canonical gauge theory of gravity (CCGG) has been applied to Friedman cosmology. CCGG is based on the covariant version of the canonical transformation theory with which all gauge theories are derived on the same footing. The difference is just the symmetry group under consideration delivering the appropriate minimal coupling scheme for matter fields and the dynamical space-time. Such a "universal" approach must of course be subject to a comprehensive testing against all kinds of experiments, especially as CCGG is its novel application to gravity. In that study \cite{Vasak:2019nmy} CCGG was shown to deliver an explanation of dark energy as a torsion based phenomenon For earlier investigations on the possible cosmological role of torsion see for example \cite{Capozziello:2002rd, Capozziello:2003tk,Capozziello:2013vna,Chen:2009at, Arcos:2005ec,Minkevich:2007eh,Minkevich:2007eh,Shie:2008ms, Unger:2018oqo}. There a first analysis of the CCGG-Friedman cosmology was limited to varying the only new parameter beyond $\Lambda$CDM. The comparison of the theory with the Hubble diagram indicated that the model can deliver viable scenarios of cosmic evolution.

In this paper that preliminary analysis is extended to Bayesian analysis with the aim to explore further cosmological constraints on the full parameter set, and to compare the results with the standard $\Lambda$CDM cosmology. After a brief review of the CCGG theory and the pertinent Friedman equations we first list the observational data considered with focus on low $z$. A discussion of the numerical analysis and the resulting figures follows. The paper concludes with a discussion of the findings.

\section{The CCGG Formulation}
\label{sec:CCGG}
Rather than following ad-hoc or trial-and-error approaches for modifying GR for compatibility with experiments, we rely ab initio on the powers of proven comprehensive mathematical frameworks. In analogy to point particle physics we apply the covariant, field theoretical version of the canonical transformation theory in the De Donder-Hamiltonian formalism to imprint a given symmetry on a system of covariant fields. In this way a consistent interaction of gravity with matter is \emph{derived} as laid out in Refs. \cite{2008IJMPE..17..435S, Struckmeier:2012dr,Struckmeier:2015vnx,Struckmeier:2017vkf, Struckmeier:2017stl,Struckmeier:2018psp}. This approach yields the \emph{Covariant Canonical Gauge Gravity} (CCGG), a Yang-Mills type theory in the spirit of Utiyama, Sciama, Kibble, Hayashi and Shirafuji, and Hehl \cite{PhysRev.101.1597,1962rdgr.book..415S,Kibble:1967sv,Hehl:1976kj,Hayashi:1981fx}, rooted in a few key assumptions. While Einstein's Principle of General Relativity translates into the requirement of diffeomorphism invariance of the coupled dynamics of matter and space-time, the Equivalence Principle is incorporated by defining at any point of space-time an inertial (observers') frame of reference. The pertinent mathematical structure is a ("Lorentzian") frame bundle with fibers spanned by ortho-normal bases fixed up to arbitrary (local) Lorentz transformations. The gauge group underlying the CCGG approach is thus the $SO(1,3)^{(+)} \times Diff(M)$ group. The emerging gauge fields are the (spin) connection coefficients not restricted to torsion-free and/or metric compatible geometries. The gauge field is a priori independent of the metric tensor, or equivalently of the vierbein fields, which come as fundamental structural elements of the Lorentzian manifold.  {Minimal couplings are discovered in that way, not postulated a priori. Of course, the structure of the dynamical space-time is dynamically implemented by a specific choice of the Hamiltonian of free gravity, but that remains the only freedom the theory leaves open. }

In order to secure the existence of the action integral in the Hamiltonian picture we also postulate \emph{non-degeneracy} of the Lagrangian and the corresponding Hamiltonian densities, which implies that the Lagrangian must contain an at least quadratic Riemann-Cartan tensor concomitant \cite{Benisty:2018ufz}. A quadratic term is therefore added as a parameter-controlled \emph{deformation} to Einstein's linear ansatz, endowing space-time with kinetic energy and thus inertia. In this way the framework delivers a classical, quadratic, first-order (Palatini) field theory where the connection coefficients emerge as independent gauge fields which, in addition to the symmetric metric tensor, determine the space-time geometry and mediate gravitation. The couplings of matter fields and gravity are unambiguously fixed. The so called consistence equation in CCGG is a combination of the canonical (or equivalently Euler-Lagrange) equations of motion, extending the field equation of GR: 
\begin{equation}
\label{eq:modEinstein}
\begin{split}
 g_1\left( R_{\alpha\beta\gamma\mu}\, {R}\indices{^{\alpha\beta\gamma}_\nu}
           - \quarter g_{\mu\nu} \, {R}_{\alpha\beta\gamma\delta}\, {R}^{\alpha\beta\gamma\delta}\right) 
           \\- \frac{1}{8\pi G}\, \left[ {R}\indices{_{(\mu\nu)}}
           - g_{\mu\nu} \left( \onehalf {R} + \lambda_0 \right)\right] = {T}\indices{_{(\mu\nu)}}.    
\end{split}
\end{equation}
Here $g_1$ is the dimensionless deformation parameter, $G$ Newton's coupling constant, and $\lambda_0$ the "bare" cosmological constant. 
\begin{equation}
R\indices{^{\alpha}_{\beta\mu\nu}} =
 \gamma\indices{^{\alpha}_{\beta\nu,\mu}} -
 \gamma\indices{^{\alpha}_{\beta\mu,\nu}} +
 \gamma\indices{^{\alpha}_{\xi\mu}}\gamma\indices{^{\xi}_{\beta\nu}} -
 \gamma\indices{^{\alpha}_{\xi\nu}}\gamma\indices{^{\xi}_{\beta\mu}}
 \end{equation}
is the Riemann-Cartan tensor (in general built from an asymmetric connection), and ${T}\indices{_{(\mu\nu)}}$ the symmetrized stress-energy tensor of matter. 
(Our conventions are the signature $(+,\,-,\,-,\,-)$ of the metric, and natural units $\hbar = c = 1$. A comma indicates a partial derivative.)

\section{The CCGG-Friedman universe}
\label{sec:CCGG-Friedman}
\subsection{Homogenous solution}
Our aim is to establish a form of the equations governing the dynamics of the universe that allow for a close comparison with GR. In particular we require that the stress-energy tensor be covariantly conserved. This requirement is here not based on the Bianchi identity for the Einstein tensor. It is invoked independently to retain the standard scaling properties of matter and radiation. As shown in Ref. \cite{Vasak:2019nmy} this leads, in a metric compatible space-time, to the necessity to invoke torsion. For the CCGG version of the Friedman model that promotes the cosmological constant to the time or scale dependent function
\begin{equation*} 
  \Lambda(a) =: \lambda_0 + \quarter P(a) =: \Lambda\,f(a).
\end{equation*}
 $P(a)$ is the torsion-dependant portion of the Ricci scalar.  $\Lambda(a)$ reduces to the "bare" cosmological constant $\lambda_0$ in torsion-free geometries. The Hubble function acquires, in addition, a further geometric correction originating from the quadratic Riemann-Cartan gravity. We ultimately get
\begin{subequations}
\label{def:rho}
\begin{equation}
E^2(a) =: \frac{H^2(a)}{H_0^2} = \frac{\rho(a)}{\rho_{crit}} \end{equation}
\begin{equation}
\rho(a) =: \rho_m + \rho_r + \rho_\Lambda + \rho_{K} + \rho_{geom} 
\end{equation}
\begin{equation}
\rho_m(a) =: \rho_{crit} \,\Omega_m \,a^{-3} 
\end{equation}
\begin{equation}
\rho_r(a) =: \rho_{crit} \,\Omega_r \,a^{-4} 
\end{equation}
\begin{equation}
\rho_{K}(a) =: \rho_{crit} \,\Omega_{K} \,a^{-3} = -\rho_{crit} \,K \,a^{-2} 
\end{equation}
\begin{equation}
\rho_\Lambda(a) =: \rho_{crit} \,\Omega_\Lambda \,f(a) = \rho_{crit} \,\onethird \Lambda\, f(a) 
\end{equation}
\begin{align}
\rho_{geom}(a) &=: \rho_{crit} \, \Omega_{\text{geom}} \\
&= \rho_{crit} \,\frac{(1/4 \Omega_m + \Omega_\Lambda)(3/4 \Omega_m + \Omega_r)}{\Omega_g - 1/4 \Omega_m -\Omega_\Lambda},
\end{align}
\end{subequations}
where $a$ is the scale factor and $K$ the curvature parameter of the FLRW metric. $\Omega_i$ with $i=m,r,\Lambda, K$ are the standard density constants related respectively to (dark) particle matter, radiation and dark energy. $H(a) = \dot{a}/a$ is the Hubble function, $H_0 \equiv H(a = 1)$ the Hubble constant, and $\rho_{crit} \equiv 3H_0^2/8 \pi G)$. For convenience we use $\Omega_g =: [32\pi G H_0^2 g_1]^{-1}$. 
We notice at this point that the various pressure terms combine to
\begin{equation}
p(a) =: \onethird\,\rho_r - \rho_\Lambda - \onethird \, \rho_{K} + \onethird \, \rho_{geom}.    
\end{equation}
Obviously the cosmological function has the equation of state $p_\Lambda = -\rho_\Lambda$ of dark energy, and the "geometric fluid" has the equation of state $p_{geom} = \onethird \, \rho_{geom}$, i.e. it behaves like (dark) radiation. 

\medskip
The normalized dark energy  function, $f(a)$, is determined from the ordinary first-order, non-linear differential equation 
\begin{equation} \label{ODE:f(a)}
f'(a) =
  \frac{3\Omega_m}{4\Omega_\Lambda \, a^4} \\\frac{
\alpha(a)
    }
  {\beta(a)
  }.     
\end{equation}
where:
\begin{equation*}
\begin{split}
 \alpha(a) =  \Omega_g \left(\threequarter \Omega_m a^{-3} + \Omega_r a^{-4}\right)\\
  - \left(\Omega_g - \quarter \Omega_m a^{-3} - \Omega_\Lambda \,f \right)
  \left(\quarter \Omega_m a^{-3} + \Omega_\Lambda \,f \right),    
\end{split}
\end{equation*}
\begin{equation*}
\begin{split}
    \beta(a) = \Omega_g \left(\threequarter \Omega_m a^{-3} + \Omega_r a^{-4}\right)  + \left(\Omega_g - \quarter \Omega_m a^{-3} - \Omega_\Lambda \,f \right)^2,    
\end{split}
\end{equation*}
with the boundary condition $f(1) = 1$. By setting $g_1 = 0$ (which means $f'(a) = 0$) and $f(a) \equiv 1$ we recover in \eref{def:rho} the Einstein-Friedman equation for the Hubble function based on General Relativity. There are five independent parameters in the CCGG model that must be optimized, namely 
$\Omega_m,\Omega_r,\Omega_\Lambda,\Omega_K$ and $\Omega_{\text{geom}} $. By solving $H_0 = H(a = 1)$, we get the relation for the additional, deformation parameter:
\begin{equation}
g_1 = \frac{1}{ 2 \pi G H_0^2 }\frac{ \Omega_K +\Omega_\Lambda +\Omega_m+\Omega_r-1}{(4 \Omega_\Lambda +\Omega_m) ( 4 \Omega_ \Lambda +\Omega_m + 4 \Omega_K - 4 )}.
\label{eq:g1}
\end{equation}
Notice that for the standard $\Lambda$CDM model (with a spatial curvature $\Omega_K$) the sum $\Omega_K +\Omega_\Lambda +\Omega_m+\Omega_r$ gives $1$ and thus from Eq. \ref{eq:g1} we get $g_1 = 0$.

 {\subsection{Stability Analysis}}
\begin{figure}[t!]
 	\centering
\includegraphics[width=0.47\textwidth]{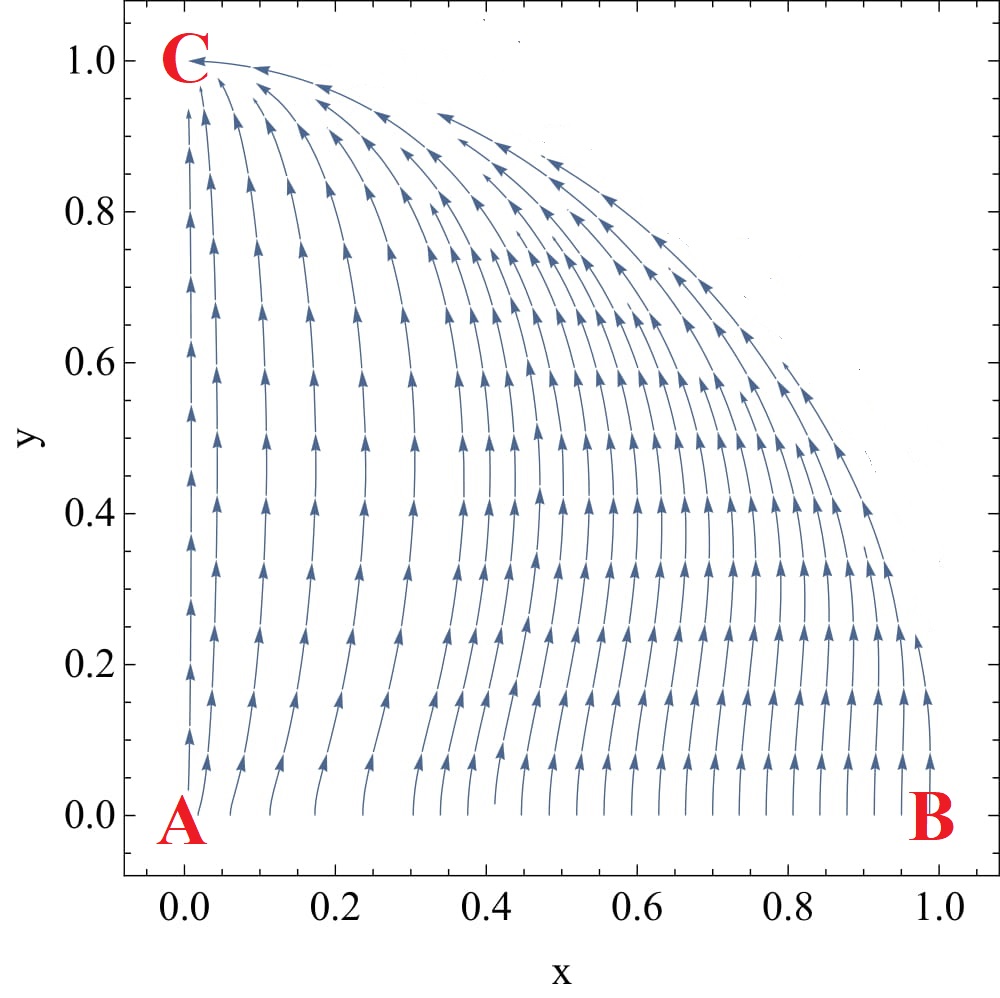}
\caption{ {\it{The stream plot for a universe with dark matter, dynamical dark energy $\Lambda(a)$ and the quadratic term from the CCGG equations. The $x$ refers to the matter part and the $y$ refers to the dark energy density. The figure shows that the matter domination (B) is an unstable point and the dark energy domination (C) is a stable point.}}}
 	\label{fig:stab}
\end{figure}

 {In order to test the stability of the model we use the autonomous system method \cite{Bahamonde:2017ize,Odintsov:2017qif}. For simplicity we ignore the spacial curvature and the radiation part and include matter, dark energy and the quadratic term that incorporates the torsional part. In that case the correct definition for the dimensionless parameters reads:}
\begin{equation}
\begin{split}
x^2=\frac{\Omega_m}{a^3 E^2},\quad y^2=\frac{\Omega_\Lambda (a)}{E^2},\quad z^2=\frac{\Omega_{geom}}{E^2}
\end{split}
\end{equation}
 {with $x^2 + y^2 + z^2 = 1$. After some algebra, one can define the evaluation equations for the system:}
\begin{subequations}
\begin{equation}
\frac{dx}{dN} = \frac{x}{2} \left(-x^2-4 y^2+1\right),
\end{equation}
\begin{equation}
\begin{split}
\frac{dy}{dN} = \frac{3 x^2 \left(x^2+y^2-1\right) \left(x^2+4 y^2-2\right)}{y \left(7 x^4+4 x^2 \left(8 y^2-5\right)+16 \left(y^2-1\right)^2\right)}\\-\frac{1}{2} y \left(x^2+4 y^2-4\right),
\end{split}
\end{equation}
\end{subequations}
 {where $N = \log(a)$. By setting $x'(N) = y'(N) = 0$, three solutions for the system are discovered. In order to estimate the stability of those points, we evaluate the matrix that contains the derivatives of the system.} 

 {Fig \ref{fig:stab} shows the stream plot for the system. Point $A (x=0,y=0)$ describes domination of the quadratic term. The eigenvalues at that point, $\lambda_{1,2} = 4, 1$, are both positive, which indicates an unstable point.
Point $B (x=1,y=0)$ where matter dominates the universe is also unstable as the eigenvalues of the point are $\lambda_{1,2} = -3,+3$. In contrast, point $C (x=0,y=1)$ with dark energy domination and the eigenvalues $\lambda_{1,2} = -3,-20$ is stable. The solution shows the evolution from point $A$ to $C$ with matter domination during the evolution of the universe.}

\section{Cosmological Probes}
\label{sec:CosProb}
\subsection{Dataset}
\begin{figure}[t!]
 	\centering
\includegraphics[width=0.47\textwidth]{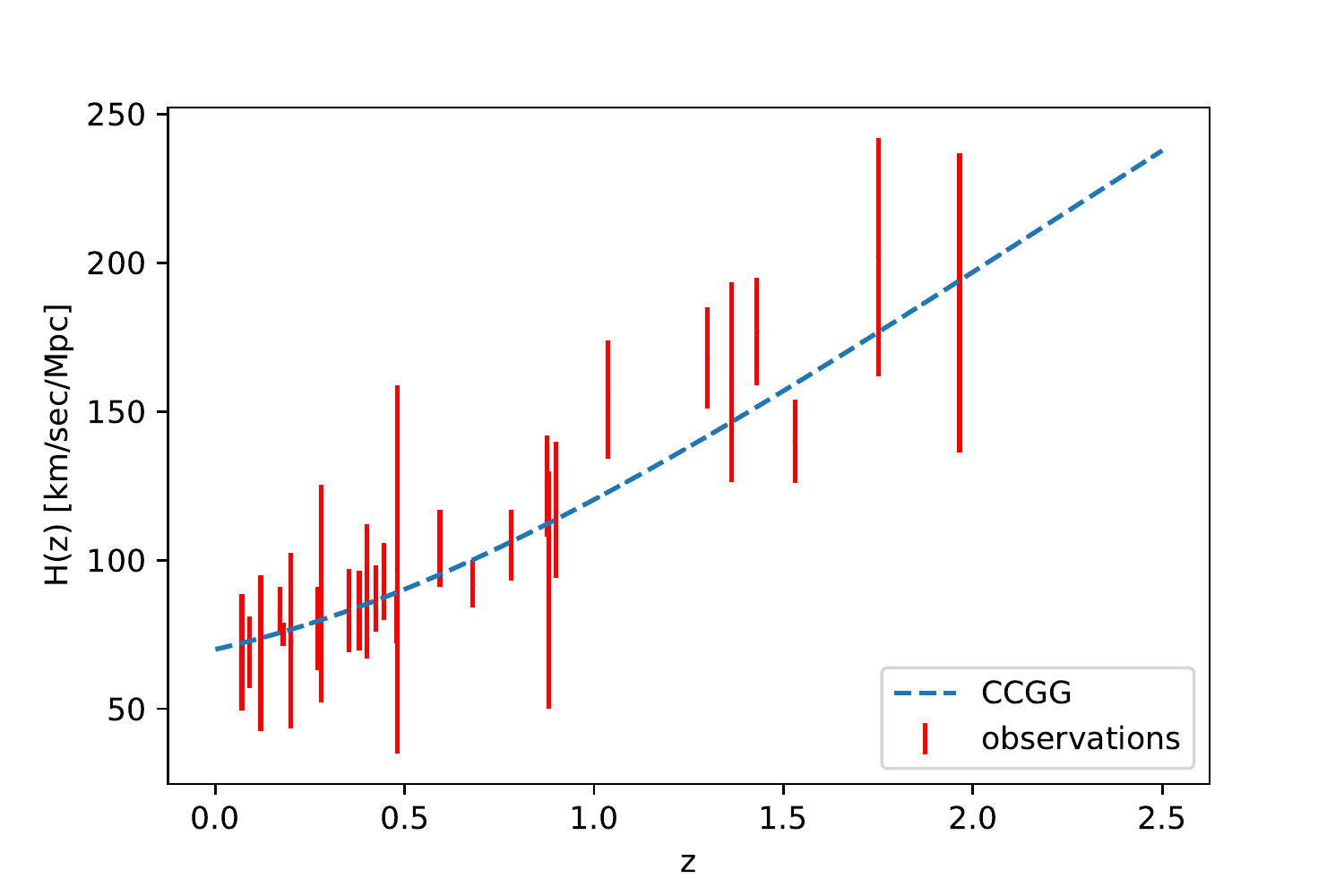}
\\
\includegraphics[width=0.47\textwidth]{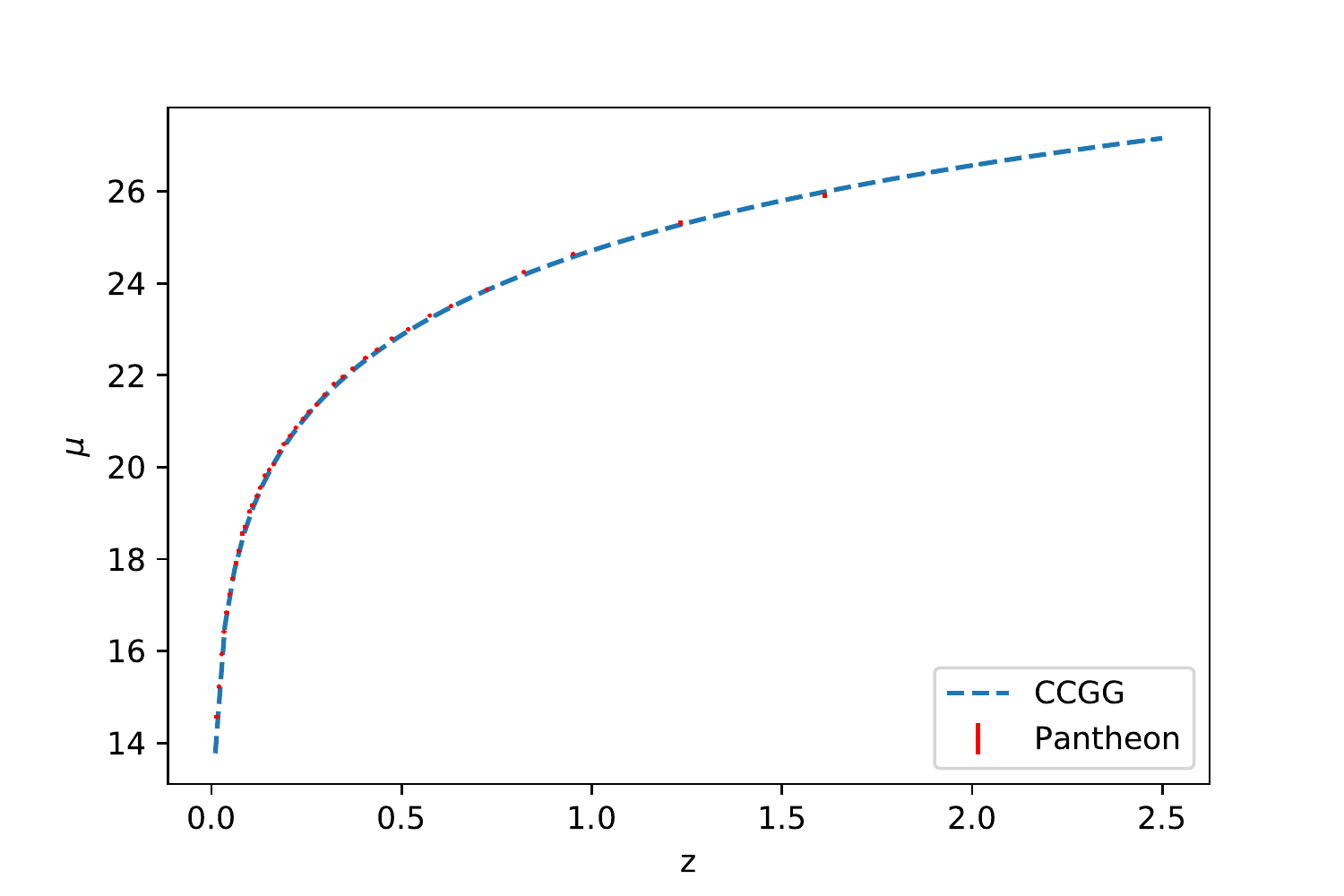}
\caption{\it{The upper panel shows the CCGG best fit vs. the Cosmic Chronometers dataset. The lower panel shows the the CCGG best fit vs. 40 uncorrelated points of the Type Ia supernova dataset. The dataset is presented in red, and the best fit is presented in blue color.}}
 	\label{fig:dataset}
\end{figure}
In order to constraint our model, we deploy the following data sets: 
\textbf{Cosmic Chronometers (CC)} exploit the evolution of differential ages of passive galaxies at different redshifts to directly constrain the Hubble
parameter \cite{Jimenez:2001gg}. We use uncorrelated 30 CC measurements of $H(z)$ discussed in \cite{Moresco:2012by,Moresco:2012jh,Moresco:2015cya,Moresco:2016mzx}. As \textbf{Standard Candles (SC)} we use uncorrelated measurements of the Pantheon Type Ia supernova \cite{Scolnic:2017caz} that were collected in \cite{Anagnostopoulos:2020ctz}. The parameters of the models are adjusted to fit the theoretical $\mu _{i}^{th}$ value of the distance modulo,
\begin{equation}
 \mu=m-M=5\log _{10}((1+z) \cdot D_{M})+\mu _{0},   
\end{equation}
to the observed $\mu _{i}^{obs}$ value.
 $m$ and $M$ are the apparent and absolute magnitudes and $\mu
_{0}=5\log \left( H_{0}^{-1}/Mpc\right) +25$ is the nuisance parameter that
has been marginalized. The luminosity distance is defined by $D_L = (1+z)\, D_M$, where 
\begin{equation}
D_M=\frac{c}{H_0} S_k\left(\int_0^z\frac{dz'}{E(z')}\right), 
\end{equation}
and 
\begin{equation}
S_k(x) = 
\begin{cases}
\frac{1}{\sqrt{-\Omega_K}}\sinh\left(\sqrt{-\Omega_K}x\right) \quad \text{if}\quad \Omega_K<0
\\
x \quad  \text{if} \quad \Omega_K=0  
\\
\frac{1}{\sqrt{\Omega_K}}\sin\left(\sqrt{\Omega_K} x\right)\quad \text{if} \quad \Omega_K>0.
\end{cases}
\end{equation}
\begin{table}[b!]
\tabcolsep 5.5pt
\vspace{1mm}
\centering
\begin{tabular}{cccc} \hline \hline
Parameter & CCGG  & CCGG + R19 & $\Lambda$CDM
\vspace{0.05cm}\\ \hline \hline
$H_0\, [\frac{km}{s\cdot Mpc}]$ & $69.3\pm 1.1$ & $71.26\pm 0.75$ & $70.18\pm 0.86$ \\
$\Omega_m$ & $0.29 \pm 0.016$ &$0.28\pm 0.014$ & $0.25\pm 0.06$ \\
$\Omega_\Lambda$ & $0.71\pm 0.09$ &$0.72\pm 0.09$ & $0.74\pm 0.04$\\
$\Omega_{r}\,{(10^{-4})} $ & $4.14\pm 2.96$ & $5.56\pm 2.87$ & - \\
$r_{d} \, \left[Mpc\right]$ & $147.4\pm 2.4$ &$143.8\pm 1.62$  & $146.1\pm 1.8$
\\
$g_1 \,(10^{114})$ & $  -0.26\pm 1.541$ &$0.0663\pm 1.094$ & 0  \\
$q$ & $-0.57\pm 0.012$ &$-0.57\pm 0.012$ & $ -0.62\pm 0.007$ \\
$\chi^2_{min}$ & $ 72.89$& $73.89$ &$74.87$
\\
$\chi^{2}_{\rm min}/Dof$& $0.92$ &$0.93$ & $0.93$\\
$AIC $& $88.51$ &$86.89$ & $84.87$\\
\hline\hline
\end{tabular}
\caption[]{\it{Observational constraints and the
corresponding $\chi^{2}_{\rm min}$ for the CCGG model with uniform prior and with the SH0ES prior, and $\Lambda$CDM model. Here we set $\Omega_K = 0$.}}

\label{tab:ResnLCDM}
\end{table}

\begin{figure*}[t!]
 	\centering
\includegraphics[width=0.45\textwidth]{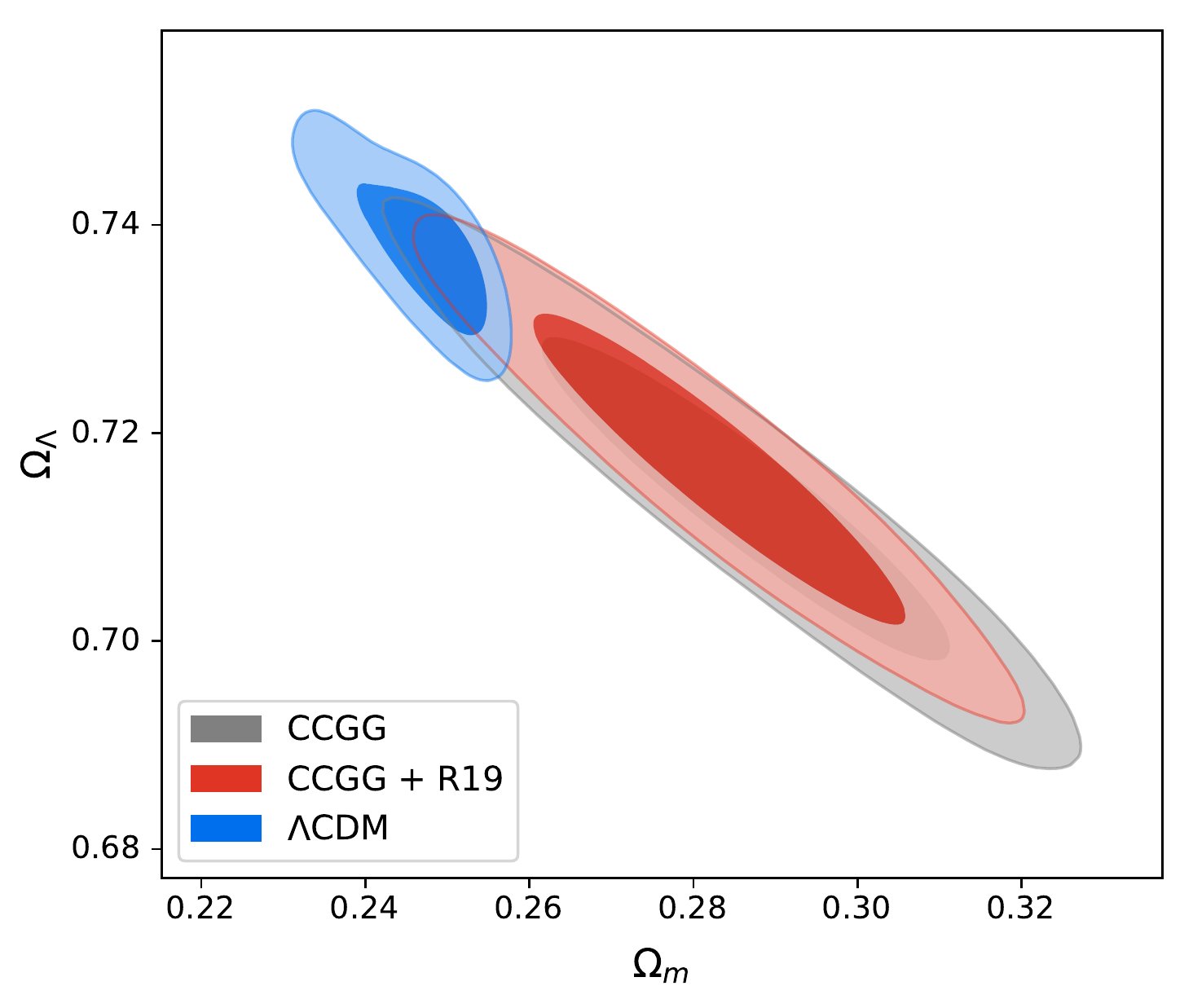}
\includegraphics[width=0.45\textwidth]{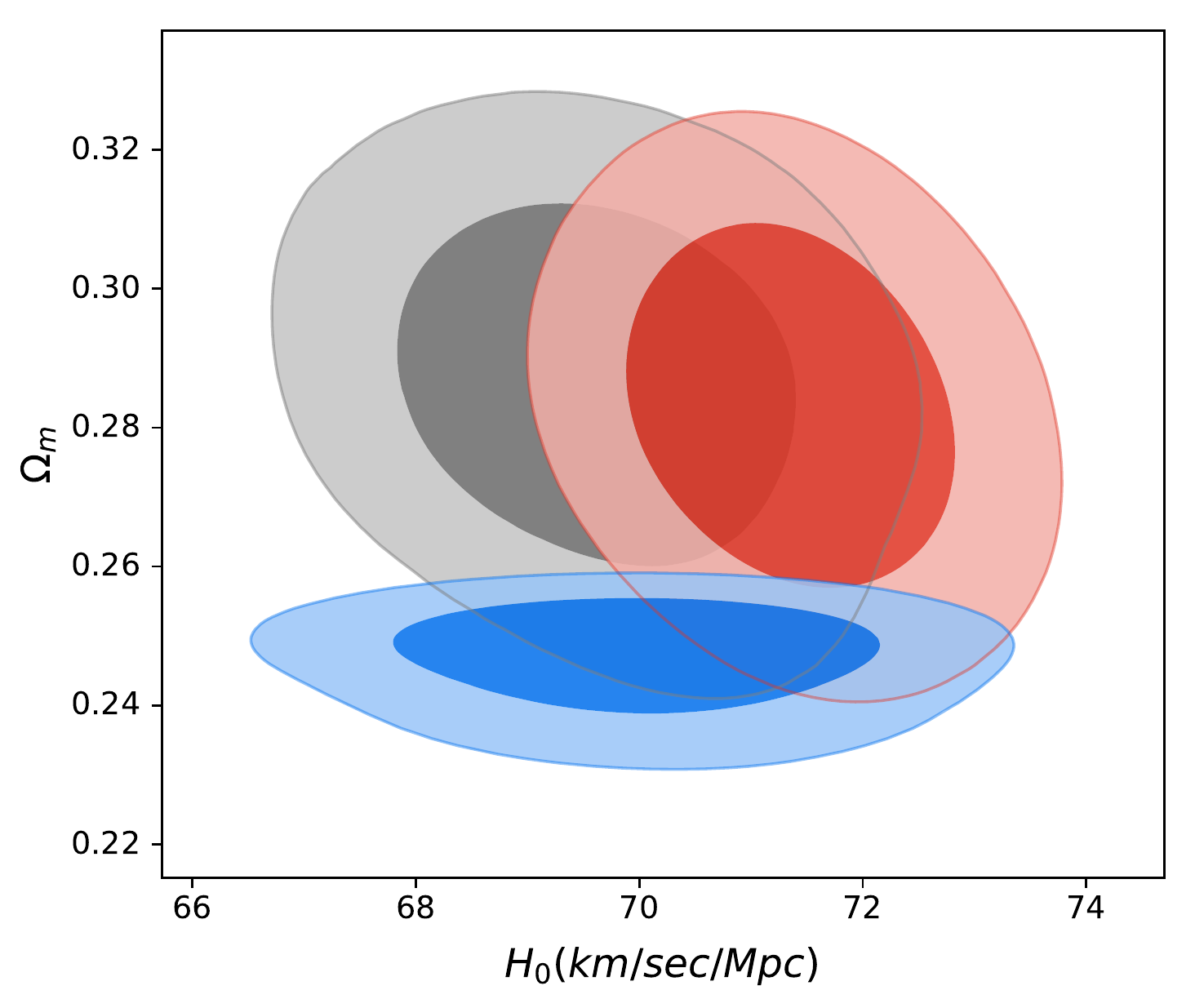}
\\
\includegraphics[width=0.45\textwidth]{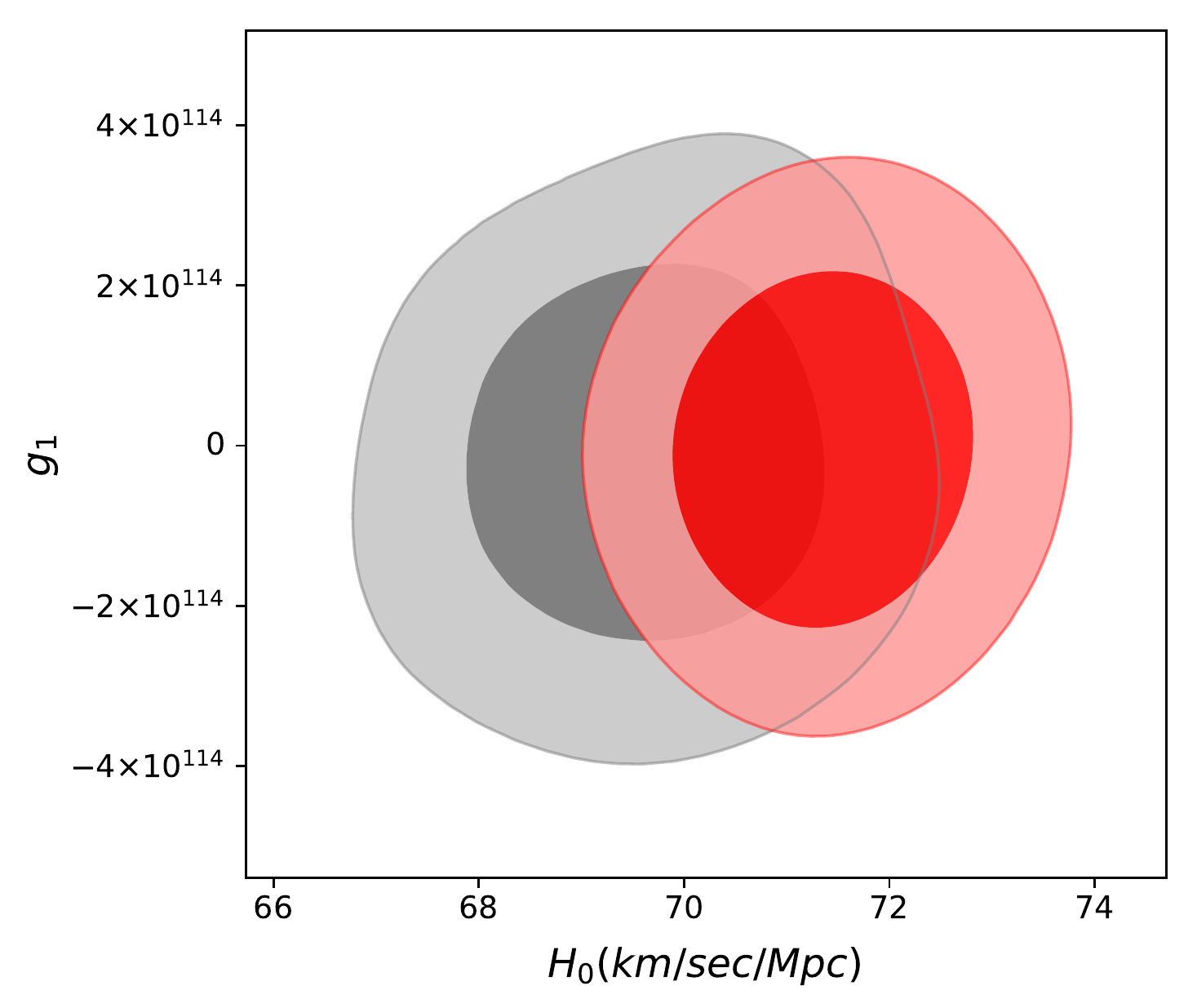}
\includegraphics[width=0.45\textwidth]{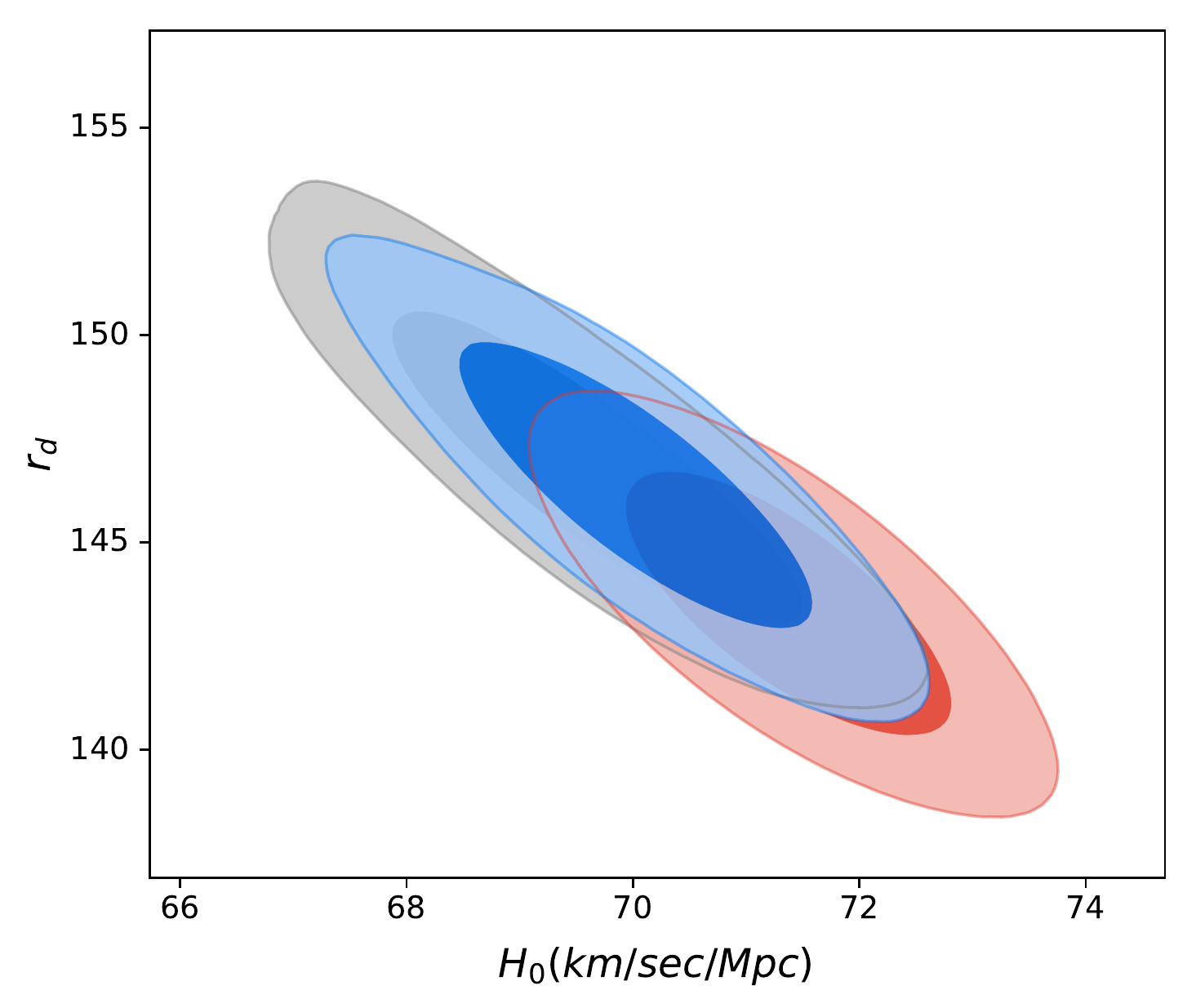}
\caption{\it{One- (68$\%$ CI) and two-dimensional (68$\%$ and 95$\%$ CI) marginalized posterior distributions for the relevant sampled and derived CCGG parameters. The upper left panel shows the contour for $\Omega_m$ vs. $\Omega_\Lambda$  and the right upper panel shows the contour for $\Omega_m$ vs. $H_0$. The lower left panel shows the contour for $g_1$ vs. $H_0$  and the lower upper panel shows the contour for $r_d$ vs. $H_0$. The gray contour describes the CCGG best fit with a uniform prior. The red contour describes the CCGG best fit with the SH0ES measurement as a prior. Finally, the blue contour describes the $\Lambda$CDM best fit with a uniform prior.}}
 	\label{fig:con}
\end{figure*}
\begin{figure}[t!]
 	\centering
\includegraphics[width=0.43\textwidth]{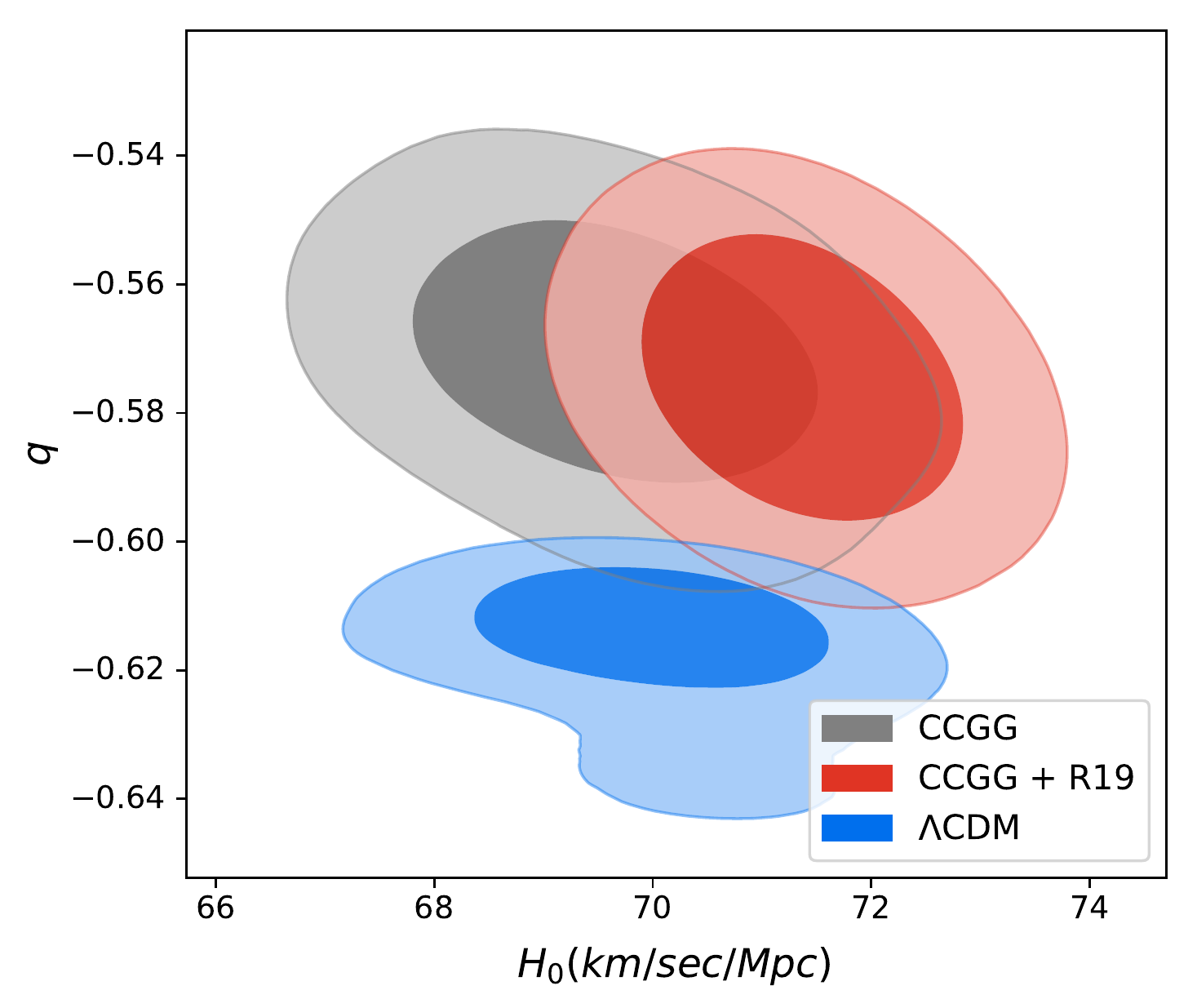}
\includegraphics[width=0.47\textwidth]{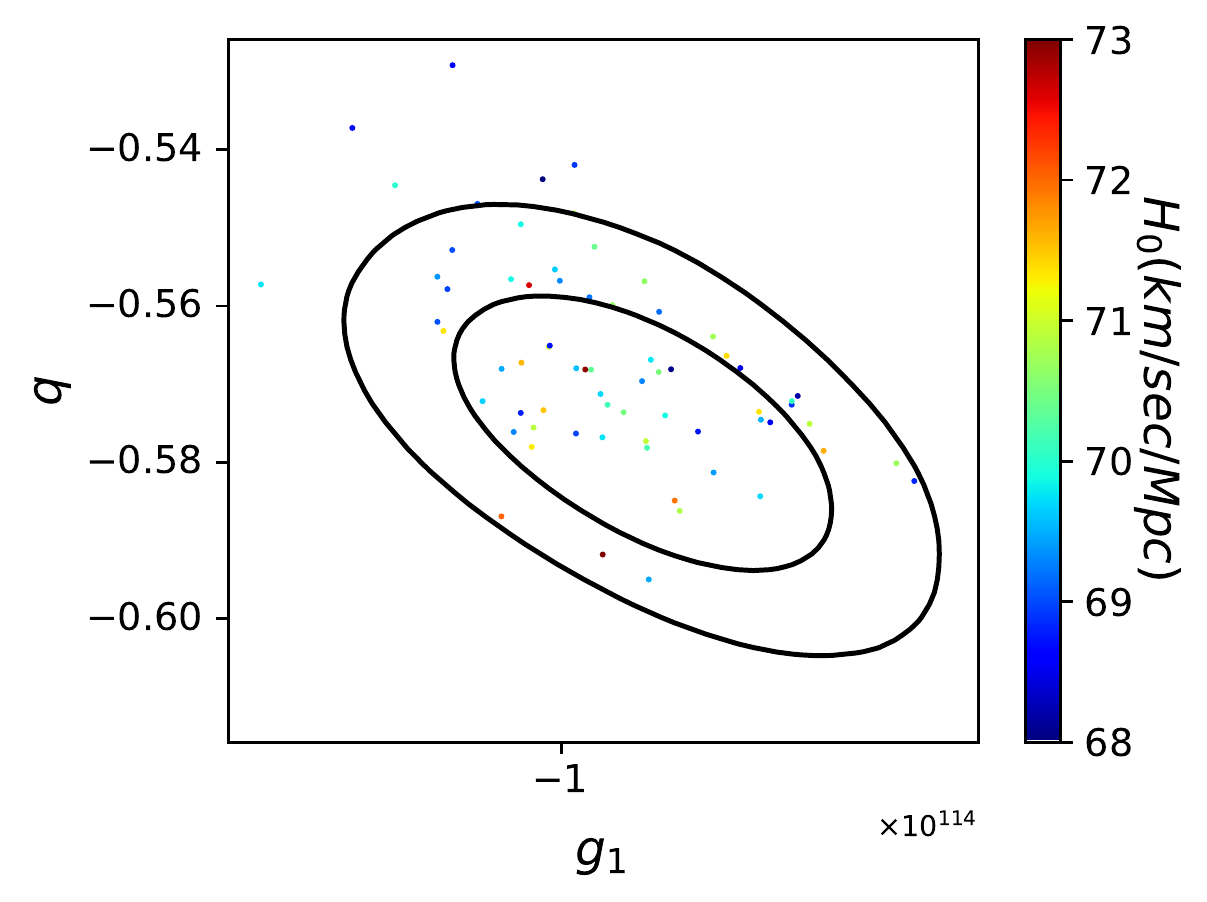}
\caption{\it{One- (68$\%$ CI) and two-dimensional (68$\%$ and 95$\%$ CI) marginalized posterior distributions for the the deceleration parameter $q$ vs. the Hubble parameter $H_0$. The gray contour describes the CCGG best fit with a uniform prior. The red contour describes the CCGG best fit with the SH0ES measurement as a prior. Finally, the blue contour describes the $\Lambda$CDM best fit with a uniform prior. The lower panel shows a 3D of $q$ and the parameter $g_1$ vs. the Hubble parameter $H_0$.}}
 	\label{fig:q}
\end{figure}
In addition, we use the uncorrelated data points from different \textbf{Baryon Acoustic Oscillations (BAO)} collected in \cite{Benisty:2020otr} from \cite{Percival:2009xn,Beutler:2011hx,Busca:2012bu,Anderson:2012sa,Seo:2012xy,Ross:2014qpa,Tojeiro:2014eea,Bautista:2017wwp,deCarvalho:2017xye,Ata:2017dya,Abbott:2017wcz,Molavi:2019mlh}. Studies of the BAO features in the transverse direction provide a measurement of $D_H(z)/r_d = c/H(z)r_d$, with the comoving angular diameter distance defined in \cite{Hogg:2020ktc,Martinelli:2020hud}. In our database we use the parameters $D_A = D_M/(1+z)$ and 
\begin{equation}
    D_V(z) \equiv [ z D_H(z) D_M^2(z) ]^{1/3}.
\end{equation}
which is a combination of the BAO peak coordinates. $r_d$ is the sound horizon at the drag epoch. Finally, for very precise  "line-of-sight" (or "radial") observations, BAO can also measure directly the Hubble parameter \cite{Benitez:2008fs}.

We use a nested sampler as it is implemented within the open-source packaged $Polychord$ \cite{Handley:2015fda} with the $GetDist$ package \cite{Lewis:2019xzd} to present the results. The prior we choose is with a uniform distributions, where $\Omega_{r} \in [0;10^{-3}]$, $\Omega_{m}\in[0.;1.]$, $\Omega_{\Lambda}\in[0.;1.]$, $H_0\in [50;100] Km/sec/Mpc$, $r_d\in [120;160] Mpc$. When we include a spatial curvature we extend the prior with $\Omega_{K} \in [-0.1;0.1]$. The measurement of the Hubble constant yielding $H_0 = 74.03 \pm 1.42 (km/s)/Mpc$ at $68\%$ CL by \cite{Riess:2019cxk} has been incorporated into our analysis as an additional prior (\textbf{R19}). We contrast best-fit parameters and goodness of fit between CCGG and the standard $\Lambda$CDM with these datasets. We also compare the Akaike information criteria (AIC) of the two models applied to the data set \cite{AIC1,Liddle:2007fy,Anagnostopoulos:2019miu}. In order to make a complete discussion, we include also the deceleration parameter in our discussion, the dimensionless quantity $q =  - 1 - \dot{H}/H^2$ measuring the acceleration of the cosmic expansion.

\begin{figure}[t!]
 	\centering
\includegraphics[width=0.45\textwidth]{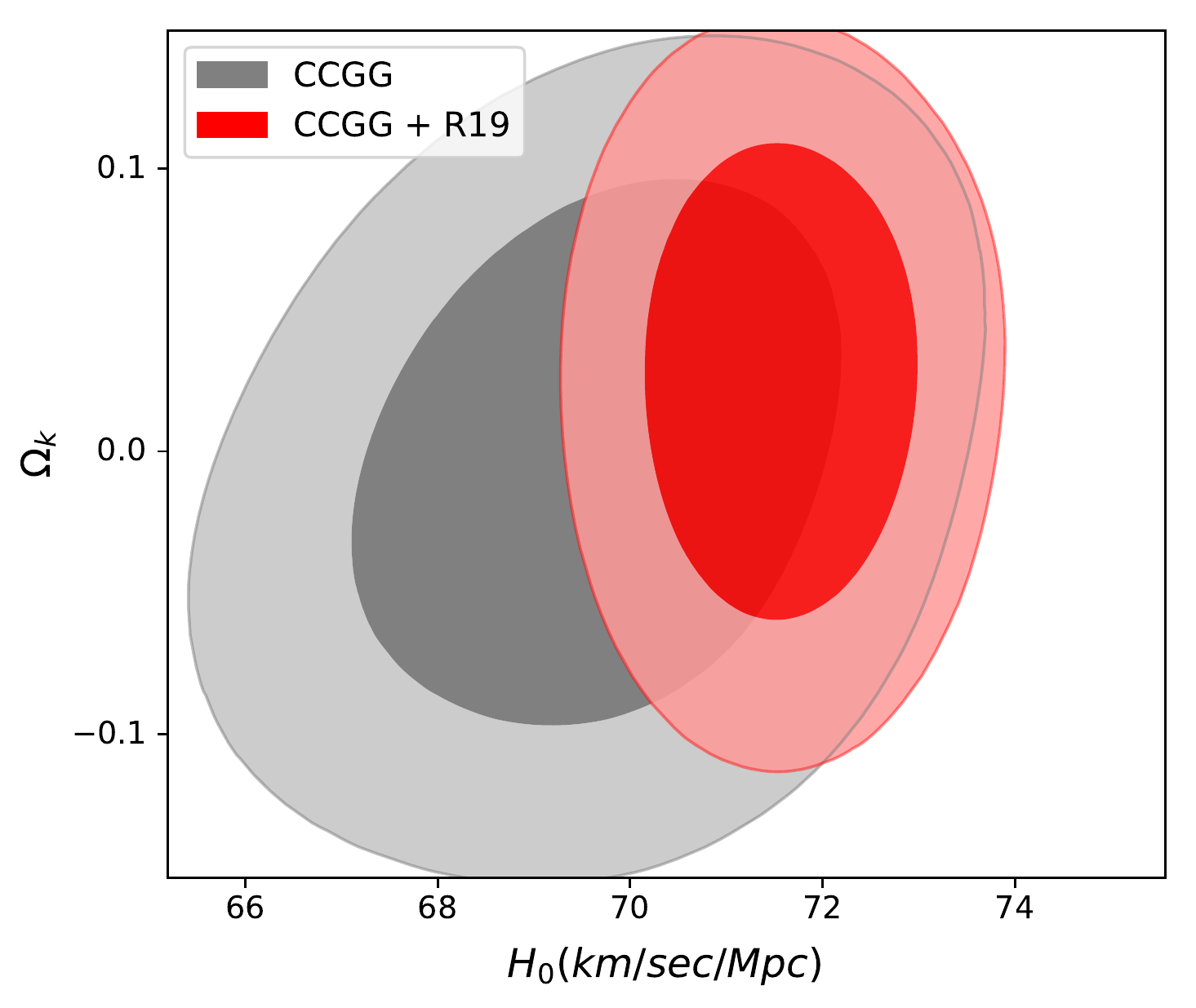}
 \begin{tabular}{ccc} \hline \hline
Parameter & CCGG  & CCGG + R19 \vspace{0.05cm}\\ \hline \hline
$H_0\, [\frac{km}{s\cdot Mpc}]$ & $69.76 \pm 1.3$ & $71.56 \pm 0.7068$  \\
$\Omega_m$ & $0.286 \pm 0.052$ &$0.26 \pm 0.042$ \\
$\Omega_\Lambda$ & $0.7141\pm 0.017$ &$0.711 \pm 0.017$ \\
$\Omega_{r}\,{(10^{-4})} $ & $5.01 \pm 2.82$ & $5.06 \pm 3.04$ \\
$\Omega_{K}\,{ (10^{-2})} $ & $0.0134 \pm 5.53$ & $2.12 \pm 4.54$\\
$r_{d} \, \left[Mpc\right]$ & $146.6 \pm 2.72$ &$143.1 \pm 1.74$  
\\
$g_1 \,(10^{114})$ & $ 0.113 \pm 2.29$ &$1.09 \pm 2.03$ \\
$q$ & $ -0.57 \pm 0.012$ &$-0.58 \pm 0.0124$ \\
$\chi^{2}_{\rm min}$& $72.83$ &$73.61$\\
$\chi^{2}_{\rm min}/Dof$& $0.93$ &$0.94$\\
$AIC $& $87.84$ &$88.61$ \\
\hline\hline
\end{tabular}
\\
\caption{\it{One- (68$\%$ CI) and two-dimensional (68$\%$ and 95$\%$ CI) marginalized posterior distributions for the the deceleration parameter $\Omega_K$ vs. the Hubble parameter $H_0$. The gray contour describes the CCGG best fit with a uniform prior. The red contour describes the CCGG best fit with the SH0ES measurement as a prior.The table shows Observational constraints and the
corresponding $\chi^{2}_{\rm min}$ for the CCGG model with uniform prior and with the SH0ES prior with spatial curvature.}}
 	\label{fig:OkH0}
\end{figure}

\subsection{Spatially flat universe}
Table \ref{tab:ResnLCDM} summarises the results with $\Omega_K = 0$. In the CCGG model the quadratic term provides with the deformation parameter an additional degree of freedom. Hence while for $\Lambda$CDM we set   $\Omega_r = 1-\Omega_m - \Omega_\Lambda$ for the radiation part, for CCGG we have  $\Omega_{\text{geom}} = 1- \Omega_m - \Omega_\Lambda - \Omega_r$ for the additional geometry term. The Hubble parameter fitted for the CCGG model is $69.3\pm 1.1 km/sec/Mpc$ for a uniform prior or $71.26\pm 0.75 km/sec/Mpc$ with the SH0ES prior. The Hubble parameter for the $\Lambda$CDM model is in between these values $70.18\pm 0.86 km/sec/Mpc$ for the uniform prior.

The $\Omega_m$ matter part in the CCGG model is  $0.29 \pm 0.016$ or $0.28\pm 0.014$ for the SH0ES prior, which is a bit higher then the $\Lambda$CDM fit $0.25\pm 0.06$. The dark energy $\Omega_\Lambda$ part is being $0.71\pm 0.09$ or  $0.72\pm 0.09$ with the SH0ES prior, a bit lower then the $\Lambda$CDM fit $0.74\pm 0.04$.

The BAO scale is set by the redshift at the drag epoch $z_d \approx 1020$ when photons and baryons decouple \cite{Aubourg:2014yra}. For a flat $\Lambda$CDM, the Planck measurements yield $147.09 \pm 0.26 Mpc$ and the WMAP fit gives $152.99 \pm 0.97 Mpc$ \cite{Aghanim:2018eyx}. Final measurements from the completed SDSS lineage of experiments in large-scale structure provide $r_d = 149.3 \pm 2.8 Mpc$ \cite{Alam:2020sor}. The $\Lambda$CDM model for the combined data set we use gives $146.1\pm 1.8 Mpc$. However, the CCGG model gives $147.4\pm 2.4 Mpc$. For the SH0ES prior, the distance is $143.8\pm 1.62 Mpc$. The quadratic term thus changes the horizon scale in the early universe, but still in a moderate and reasonable range.

From the AIC we see that $\Lambda$CDM is still the better fit to the late universe, since the AIC for $\Lambda$CDM model $84.87$ is then the CCGG case  $88.51$ or with SH0ES prior $86.89$. However the $\Lambda$CDM model does not describe the inflationary epoch, which the quadratic term naturally provides.

\subsection{With $\Omega_K \ne 0$}
The shape of the universe is a fundamental question. The latter can be characterized by measuring the \textit{spatial curvature} of the universe $K$, quantifying how much the spatial geometry locally differs from that of flat space. Most models of inflation predict a universe which is extremely close to being spatially flat~\cite{Baumann:2009ds,Martin:2013tda,Aghanim:2018eyx}. Because of the quadratic term the spatial curvature may be larger. Fig \ref{fig:OkH0} shows the spatial curvature vs. the Hubble parameter. For the uniform prior case the spatial curvature turns out to be $\left(0.0134 \pm 5.53\right)\cdot 10^{-2}$, while for the SH0ES prior the spatial curvature is $\left(2.12 \pm 4.54 \right)\cdot 10^{-2}$. The model predicts a positive value for $\Omega_K$ but the error bar is sufficient large for the negative values as well. Moreover, from the AIC criteria it seems that the case for absorbing the spatial curvature is better since the AIC for this case is higher: $87.84$ for the uniform prior, and $88.61$  for the SH0ES prior.

\section{Discussion}
\label{sec:Dis}
This paper discusses the cosmological constraints on the CCGG formulation from low-redshift observations. CCGG is a gauge theory of gravity ensuring  in a covariant way full diffeomorphism invariance of the system action. Using canonical transformation theory the approach unambiguously fixes how matter fields interact with curved geometry of space-time, and enforces a parameter controlled admixture of a quadratic Riemann-Cartan concomitant to the Einstein-Hilbert linear term. In a preliminary study \cite{Vasak:2019nmy} the cosmological consequence of that quadratic extension were examined in alignment with the $\Lambda$CDM model. Here we go a step further and test  the CCGG cosmology against a comprehensive  database of low-redshift cosmological measurements that include the Pantheon Type Ia supernova, Cosmic Chronometers and Baryon Acoustic Oscillations. 

Using the Polychord package we find a good best fit of the CCGG cosmology  with data. By the AIC criterium the CCGG fit accuracy is comparable with that of $\Lambda$CDM.  The key density parameters are in reasonable agreement with the $\Lambda$CDM model. The new free parameter of the theory controlling the admixture of quadratic gravity and the inflationary dynamics is of the order $10^{114}$. However, the statistical error bars do not permit any further conclusions about its value, not even about its sign. 
Also the non-zero  spatial curvature parameter found for the late universe is, within the error bars, consistent with zero. The deceleration parameter $q$, on the other hand, is predicted to be lower then the $\Lambda$CDM best fit: $-0.57\pm 0.012$ for CCGG and $ -0.62\pm 0.007$ for $\Lambda$CDM.

\medskip
We conclude that the CCGG approach reproduces low-redshift observation with a similar accuracy as the $\Lambda$CDM model. The novel features of CCGG, namely the presence of torsion and the influence of quadratic curvature, represented by the additional \emph{deformation} parameter, turn out to be  subdominant in this late era of the cosmic evolution. However, albeit this calculation does not conclusively determine the relative  admixture of quadratic gravity to Einsten-Cartan gravity, the data neither excludes a non-zero  deformation parameter, nor a deviation from the flat geometry assumed in  $\Lambda$CDM. This leaves the possibility open that the more complex space-time geometry of CCGG applies, which naturally invokes inflation \cite{Benisty:2018ywz} and substantially alters the Hubble expansion in the early universe \cite{Vasak:2019nmy}. The model's superiority thus might become obvious only when including the early universe data in the analysis. 

 {The discussion on whether torsion of space-time could be excluded by solar tests have been sparked by the Gravity Probe B (GPB) experiment. While Mao et al. \cite{Mao:2006bb} propose to use the high-precision gyroscope for detecting torsion, Hehl et al. \cite{Hehl:2013qga} conclude, referring to the Poincare Gauge Gravity, that  torsion can couple to particle spin only, not to the gyroscope's angular momentum, and that the accuracy needed for detecting such a coupling is far beyond any currently available technologies. Moreover, they also dismiss possible deviations of test particle trajectories from the geodesic (trajectory of extremal length) by postulating that it is force-free trajectory of particles, rather than the autoparallel ("straightes trajectory"). These discussions indicate that torsion may not be detectable in the solar system, but they so far do not exclude its existence.}  

 {While we agree with Hehl's first conjecture, we advocate the autoparallel to be the correct force-free trajectory as the obvious  generalization of Newton's notion of a straight line to curvilinear space-time. A propagating torsion field that arises naturally in CCGG will also directly interact with spin carrying particles. As in presence of torsion the autoparallel and the geodesic are not identical, a modified connection will in principle be felt by test particles. Direct spin-torsion interactions \cite{Struckmeier:2021rst} will in addition affect the trajectories of spin-polarized as compared to spinless test particles. However, if the density of the torsion field is very low in the solar system then we will encounter similar restrictions on the detectability of those spin-torsion interactions. } Work along these lines including advanced modeling of the torsion tensor is in progress.   

 {\textbf{Public Source}: The files with the dataset and the fit package can be found in \url{https://github.com/benidav/CCGGcosmology2020}{}.}

\acknowledgements
We thank Horst Stöcker and Denitsa Sticova for fruitful discussions. This work has been supported by the Walter Greiner Gesellschaft zur Förderung der physikalischen Grundlagenforschung e.V., and partially by the European COST actions CA15117 and CA18108. D.B., D.V. and J.K. especially thank the Fueck Stiftung for support.

\bibliographystyle{apsrev4-1}
\bibliography{ref}

\end{document}